\documentclass[12pt]{article} 
\usepackage{epsf} 
\usepackage{epsfig} 
 
\newcommand{\sss}{\scriptscriptstyle}

\def\bc{\begin{center}}
\def\ec{\end{center}}
\def\be{\begin{equation}}
\def\ee{\end{equation}}
\def\bea{\begin{eqnarray}}
\def\eea{\end{eqnarray}}
\def\nn{\nonumber}

\begin{document}
\pagestyle{empty} 
\vspace{-0.6in}
\begin{flushright}
RM3-TH/01-3 \\
ROMA-1310/01 
\end{flushright}
\vskip 1.2 cm
\centerline{\Large{\bf{CHARMING PENGUINS STRIKE BACK}}}
\vskip 1.4cm
\centerline{{\bf M.~Ciuchini$^{1}$, E.~Franco$^2$,
G. Martinelli$^{2}$, M.~Pierini$^2$,  L.~Silvestrini$^2$}}
\vskip 1.cm
\centerline{$^1$   Dip. di Fisica, Univ. di Roma Tre 
and INFN, Sezione di Roma Tre,}
\centerline{Via della Vasca Navale 84, I-00146   
Roma, Italy}
\centerline{$^2$ Dip. di Fisica, Univ. ``La Sapienza''  and INFN,}
\centerline{Sezione di Roma, P.le A. Moro, I-00185 Rome, Italy.}
\vskip 1.4cm \abstract{By using the recent experimental measurements
of $B \to \pi\pi$ and $B \to K \pi$ branching ratios, we find that
factorization is unable to reproduce the observed $BR$s even taking
into account the uncertainties of the input parameters. Charming and
GIM penguins allow to reconcile the theoretical predictions with the
data.  Because of these large effects, we conclude, however, that it
is not possible, with the present theoretical and experimental
accuracy, to determine the CP violation angle $\gamma$ from these
decays.  Contrary to factorization, we predict large asymmetries for
several of the particle--antiparticle $BR$s, in particular $BR(B^{+}
\to K^+ \pi^0) $, $BR(B_d \to K^+ \pi^-) $ and $BR(B_d \to \pi^+
\pi^-)$. This opens new perspectives for the study of CP violation in
$B$ systems.  }
\vskip 5 cm
\vfill\eject
\pagestyle{empty}\clearpage
\setcounter{page}{1}
\pagestyle{plain}
\newpage 
\pagestyle{plain} \setcounter{page}{1}
\section{Introduction} \label{intro}
The theoretical understanding of non-leptonic two body $B$ decays is a
fundamental step for testing flavour physics and CP violation in the
Standard Model and for detecting signals of new
physics~\cite{Ciuchini:1997zp}--\cite{Baek:1999yn}.  The increasing
accuracy of the experimental measurements at the $B$
factories~\cite{babar,belle} calls for a significant improvement of
the theoretical predictions.  In this respect,  important progress
has been recently achieved by systematic studies of factorization made
by two independent groups~\cite{li,beneke}. These studies, while
confirming the physical idea~\cite{previous} that factorization holds
for hadrons containing heavy quarks, $m_{Q} \gg \Lambda_{QCD}$, give
the explicit formulae necessary to compute quantitatively the relevant
amplitudes at the leading order of the  $\Lambda_{QCD}/m_{Q}$ expansion.  They also
examine some of the contributions entering at higher order in
$\Lambda_{QCD}/m_{Q}$.  The question which naturally arises
is whether in practice the power-suppressed corrections, for which
quantitative estimates are missing to date, may be phenomenologically
important for $B$ decays.  This problem was previously addressed in
refs.~\cite{charming, pham, fleischer}. In particular, the main
conclusion of refs.~\cite{charming} was that {\it non-perturbative}
penguin contractions of the leading operators of the effective weak
Hamiltonian, $Q_{1}$ and $Q_{2}$, although formally of ${\cal
  O}(\Lambda_{QCD}/m_{Q})$, may be important in cases where the
factorized amplitudes are either colour or Cabibbo suppressed.  The
most dramatic effect of these non-factorizable penguin contractions
manifested itself in the very large enhancement of the $B \to K \pi$
branching ratios, as was also emerging from the first measurements by
the CLEO Collaboration~\cite{cleo}.  In this case, the effect was
triggered by Cabibbo-enhanced penguin contractions of the operators
$Q^{c}_{1}$ and $Q^{c}_{2}$, usually referred to as {\it charming
  penguins}. Since the original publications, about three years ago,
several other decay channels have been
measured~\cite{cleobr,babarbr,bellebr} and the precision of the
measurements is constantly improving.  With respect to 
previous analyses, it is now possible to attempt a more quantitative
study of charming penguin effects and of the corrections expected to
the factorized predictions.  We now present the main conclusions of
our new analysis.
\subsubsection*{Factorization with $\vert  V_{ub}\vert$  and $\gamma$ 
  from other determinations} 

Using the available experimental information on $\vert V_{ub}\vert$
and on the CP angle $\gamma$ provided by the unitarity triangle
analysis (UTA)~\cite{Ciuchini:2000de}, the branching ratios predicted
with the factorized amplitudes, including the ${\cal O}(\alpha_{s})$
corrections computed according to ref.~\cite{beneke}, fail to
reproduce the experimental $B \to K \pi$  branching ratios
that  are systematically larger than the theoretical predictions.  In
addition, $BR(B_{d}\to \pi^{+} \pi^{-})$, which depends on the
semileptonic form factor $f_{\pi}(0)$, is about a factor of 2 larger
than its experimental value~\footnote{ Unless explicitly stated the
  $BR$s always refer to the average of particles and antiparticles,
  e.g.  $BR(B_{d} \to K^{0} \pi^{0})\equiv 1/2 ( BR(\bar B^{0}_{d} \to
  \bar K^{0} \pi^{0})+ BR(B^{0}_{d} \to K^{0} \pi^{0}))$.}. We note
that the value of $BR(B_{d} \to \pi^{+}\pi^{-})$ within factorization
is essentially fixed by the measured $BR(B^{+} \to \pi^{+}\pi^{0})$
rate. Thus, contrary to the statement of ref.~\cite{beneke}, the
predicted value of $BR(B_{d} \to \pi^{+}\pi^{-})$ is independent of
the theoretical assumptions on the value of $f_{\pi}(0)$. This holds
essentially true also for the $B \to K\pi$ $BR$s since the value of
the ``semileptonic'' form factor at zero momentum transfer $f_{K}(0)$
is correlated to $f_{\pi}(0)$ by the approximate $SU(3)$ symmetry.
\subsubsection*{Factorization fitting $\gamma$}  
Even if one ignores the value of $\gamma$ from UTA, which is only
justified if there are contributions to $\Delta F=2$ mixing due to
physics beyond the Standard Model, there are serious difficulties in
reproducing the experimental results. In particular, $BR(B_{d} \to
K^{0} \pi^{0})$ and $BR(B^{+} \to K^{0} \pi^{+})$ are much smaller
than their experimental values.  Moreover, the value of $\gamma$
extracted from a fit to the data, $\gamma
= (163 \pm 12)^{o}$, is in total disagreement with that from the UTA,
$\gamma=(54.8 \pm 6.2)^{o}$~\cite{Ciuchini:2000de}.  In addition, in
order to enhance the $B \to K \pi$ rates, the preferred values of
$f_{K}(0)=0.40\pm 0.02$ and $f_{\pi}(0)=0.34\pm 0.01$ are incompatible
with the latest theoretical estimates, $f_{\pi}(0)= 0.26\pm 0.05\pm
0.04$, $f_{K}(0)/f_{\pi}(0)=1.21 \pm 0.09
^{+0.00}_{-0.09}$~\cite{latticeff} and $f_{\pi}(0)=0.28\pm 0.05$,
$f_{K}(0)/f_{\pi}(0)=1.28^{+0.18}_{-0.10}$~\cite{qcdsrff}, whereas
$\vert V_{ub}\vert$ must have a rather low value, $\vert
V_{ub}\vert=(2.79 \pm 0.19) \times 10^{-3}$ instead of that extracted
from inclusive~\cite{inclusivevub} and exclusive~\cite{exclusivevub}
semileptonic $B$ decays, $\vert V_{ub}\vert=(3.25 \pm 0.29 \pm 0.55)
\times 10^{-3}$. We conclude that, even relaxing the constraint on
$\gamma$, it is very difficult to reconcile the predictions from
factorization with the experimental and theoretical findings.  For
this reason any attempt to extract, within factorization, the value of
$\gamma$ from ratios of $BR$s, for which the discrepancies with the
experiments can be accidentally hidden, is not very useful. We think
that a preliminary step is to understand the missing dynamical
effects.

\subsubsection*{Factorization and charming penguins}  

The inclusion of charming penguin effects, which will be explained in
detail in sec.~\ref{sec:res}, considerably improves the situation for
the $B \to K \pi$ channels, with values of $\vert V_{ub}\vert$ and
$\gamma$ well compatible with other determinations. In contrast to the
$B \to K \pi$ case, charming penguins are not Cabibbo enhanced in $B
\to \pi \pi$ decays and are thus expected {\it a priori} to play a
minor role. For this reason they should be consistently neglected,
together with all other $\Lambda_{QCD}/m_{b}$ corrections.  This would
leave the problem of a too large predicted $BR(B_{d} \to
\pi^{+}\pi^{-})$ unsolved.   A natural question is then whether 
the inclusion of  $\Lambda_{QCD}/m_{b}$ effects in $B_{d} \to
\pi^{+}\pi^{-}$  can  improve the agreement of the predictions with the
experimental data. 
In particular, besides the charming penguins,
penguin contractions of $Q_{1}^{u}$ and $Q_{2}^{u}$
(GIM penguins in the notation of ref.~\cite{charming}), which are
Cabibbo suppressed in $B \to K \pi$, might play an important r\^ole. 
We show that, for numerical values of the charming and GIM
penguin amplitudes of the expected size, $\Lambda_{QCD}/m_{b}\sim
0.1$--$0.2$, we can easily reproduce the experimental data for both $B
\to K \pi$ and $B \to \pi \pi$ decays while respecting the constraints
from the UTA. The sensitivity of $B \to \pi^{+} \pi^{-}$ to
$\Lambda/m_{b}$ effects casts serious doubts on the possibility of
extracting $\sin 2 \alpha$ from the coefficient of the $\sin \Delta
m_{B_d} t$ term obtained from CP asymmetry measurements. On the other
hand, we find that the value of the rate asymmetry,
\begin{equation}
  {\cal A}(B_d \to \pi^+ \pi^-) = \frac{BR(\bar B^{0}_{d} \to
    \pi^{+}\pi^{-})-BR(B^{0}_{d} \to \pi^{+}\pi^{-})}{BR(\bar
    B^{0}_{d} \to \pi^{+}\pi^{-})+BR( B^{0}_{d} \to \pi^{+}\pi^{-})}
  \, ,
\label{eq:asy} 
\end{equation} 
could be unexpectedly large and call our experimental colleagues for
separate measurements of the $B$ and $\bar B$ $BR$s.
In particular, we find $\vert {\cal A}(B^\pm \to K^\pm \pi^0)\vert =
0.18 \pm 0.06$, $\vert {\cal A}(B_d \to K^\pm \pi^\mp)\vert = 0.17 \pm
0.06$.  We also find $\vert {\cal A}(B_d \to \pi^+ \pi^-)\vert=
0.30$--$0.50$. In the latter case, as discussed in the following, the
results are subject to other effects on which we do not have control.
For this reason we do not quote an error. We simply signal that there
is room for a large asymmetry also in $B_d \to \pi^+ \pi^-$ decays.
\section{Results}
\label{sec:res}

In this section we describe and discuss more in detail the different
cases which have been considered in our analysis.

The physical amplitudes for $B \to K \pi$ and $B \to \pi \pi$ are more
conveniently written in terms of RG invariant parameters built using
the Wick contractions of the effective Hamiltonian~\cite{BS}.  In the
heavy quark limit, following the approach of ref.~\cite{beneke}, it is
possible to compute these RG invariant parameters using factorization.
The formalism has been developed so that it is possible to include
also the perturbative corrections to order $\alpha_{s}$, i.e. at the
next-to-leading order in perturbation theory.  We present results
obtained with this formalism with the addition of
the non-perturbative $\Lambda_{QCD}/m_{b}$ corrections to
factorization described below in this section.  An alternative
framework is provided by the approach of ref.~\cite{li}.  This method
differs in the treatment of the ${\cal O}(\alpha_{s})$ terms; unlike
the method of ref.~\cite{beneke}, the calculations are only valid at
the leading logarithmic order and it is not clear how the independence
of the final result from the renormalization scale of the operators of
the effective Hamiltonian is recovered.  Moreover the Sudakov
suppression of the endpoint region, advocated in~\cite{li}, is still
rather controversial from both the theoretical and phenomenological
point of view. For these reasons we prefer to postpone the analysis
with the approach of ref.~\cite{li} until the theoretical situation will
become clearer.

In the leading amplitudes, we have taken into account the SU(3)
breaking terms by using the appropriate decay constants, $f_K$ and
$f_\pi$, and form factors, $f_K(0)$ and $f_\pi(0)$. Strictly speaking,
the form factors should be evaluated at the invariant mass of the
emitted meson ($f_K(m_\pi^2)$, $f_\pi(m_K^2)$ or $f_\pi(m_\pi^2)$).
The difference is however of higher order in $\Lambda_{QCD}/m_{b}$ and
not Cabibbo or colour enhanced and can safely be neglected (it is also
numerically immaterial)~\cite{cottingam}.  As for
$\Lambda_{QCD}/m_{b}$ corrections, we have assumed instead $SU(3)$
symmetry and neglected Zweig-suppressed contributions.  In this
approximation, by $SU(3)$ symmetry one can show that all the
Cabibbo-enhanced $\Lambda_{QCD}/m_{b}$ corrections to $B \to K \pi$
decays can be reabsorbed in a single parameter $\tilde P_{1}$.
Several corrections are contained in $\tilde P_{1}$: this parameter
includes not only the charming penguin contributions, but also
annihilation and penguin contractions of penguin operators. It does
not include leading emission amplitudes of penguin operators
($Q_3$--$Q_6$) which have been explicitly evaluated using
factorization.  Had we included these terms, this contribution would
exactly correspond to the parameter $P_1$ of ref.~\cite{BS}.  The
parameter $\tilde P_{1}$ ($P_{1}$) encodes automatically not only the
effect of the annihilation diagrams considered in~\cite{keum}, but
all the other contributions of ${\cal O}(\Lambda_{QCD}/m_{b})$ with
the same quantum numbers of the charming penguins. In this respect it is
the most general parameterization of all the perturbative and
non-perturbative contributions of the operators $Q_{5}$ and $Q_{6}$
($Q_{3}$ and $Q_{4}$), including the worrying higher-twist infrared
divergent contribution to annihilation discussed in
ref.~\cite{beneke2}.  The parameter $\tilde P_{1}$ is of ${\cal
  O}(\Lambda_{QCD}/m_{b})$ and has
the same quantum numbers and physical effects
as the original charming penguins proposed in~\cite{charming}, 
although it has a more general meaning.  In some of the
previous analyses, see for example~\cite{ali}, penguin contractions of
the operator $Q_6$, computed by using perturbation theory and
factorization, are enhanced by taking a low effective scale for
$\alpha_s$.  This procedure produces a physical effect similar to that
coming from the non-perturbative charming penguins that we are using
here, since they have the same quantum numbers. 

If one also includes $B \to \pi\pi$ decays we have several other
parameters, for example $P_{1}^{\sss {\rm GIM}}$ and $P_{3}$, in the formalism of
ref.~\cite{BS}. A closer look to $P_{3}$ shows that this term is due
either to Zweig suppressed annihilation diagrams (called CPA and DPA
in ref.~\cite{charming}) or to annihilation diagrams which are colour
suppressed with respect to those entering $\tilde P_{1}$. For
this reason we have put $P_{3}$ to zero.  $P_{1}^{\sss {\rm GIM}}$ will be
discussed later on.
 
We give now the explicit expression of the $B_d \to K^+ \pi^-$
amplitude as an illustrative example. In terms of the parameters
defined in~\cite{BS}, this amplitude reads 
\begin{eqnarray} {\cal
    A}(B_d \to K^+ \pi^-) = &-& V_{us} V_{ub}^*
  \Bigl(E_1(s,u,u;B_d,K^+,\pi^-) - P_1^{\sss {\rm
      GIM}}(s,u;B_d,K^+,\pi^-)\Bigr)\nn \\ 
&+& V_{ts} V_{tb}^*
  \,P_1(s,u;B_d,K^+,\pi^-)\,.  
\end{eqnarray} 
Using the approach
of~\cite{beneke}, we have 
\begin{eqnarray}
  E_1(s,u,u;B_d,K^+,\pi^-)&=& a_1^u(K \pi) \langle Q_1^u\rangle_{\rm
    fact} + a_2^u(K \pi) \langle
  Q_2^u\rangle_{\rm fact} + \tilde E_1\nn \\
  P_1(s,u;B_d,K^+,\pi^-)&=&\sum_{i=3}^6 a_i^c(K\pi) \langle Q_i
  \rangle_{\rm fact} +\tilde P_1 \nn \\ 
P_1^{\sss {\rm
      GIM}}(s,u;B_d,K^+,\pi^-)&=& \sum_{i=3}^6
  (a_i^c(K\pi)-a_i^u(K\pi)) \langle Q_i \rangle_{\rm fact} +\tilde
  P_1^{\sss {\rm GIM}}\, , 
\end{eqnarray} where $\langle Q_i
\rangle_{\rm fact}$ denotes the factorized matrix element, and the
parameters $a_i$ are defined in~\cite{beneke}. The tilded parameters
represent $\Lambda_{QCD}/m_{b}$ corrections; in $B \to K \pi$ channels
the only Cabibbo-enhanced correction is given by $\tilde P_1$. This
term has no arguments since we take it in the $SU(3)$ symmetry limit.

We use input parameters (like $\bar \rho$,
$\bar \eta$, the form factors) with errors, and extract output
quantities (like the $BR$s, the asymmetries, but also $\gamma$, or the
form factors when they are not used as inputs) with their
uncertainties.  Let us explain how we used the input errors
and extracted the output uncertainties. We proceed with the usual
likelihood method, by generating the input quantities weighted by
their probability density function (p.d.f.). In the case of theoretical 
quantities this is
assumed to be flat, whereas the experimental quantities are extracted
with Gaussian distributions.  Probability density functions,
averages and standard deviations are then obtained by weighting the
output quantities by the likelihood factor 
\begin{equation} {\cal L} =
  e^{- \frac{1}{2} \sum_{i} {(BR_{i} -BR_{i}^{exp})^{2}}/{
      \sigma_{i}^{2}}} \, ,
\end{equation} where $\sigma_{i}$ are the
standard deviations of the experimental $BR$s, $BR^{exp}_{i}$, given in
table~\ref{tab:inputs}. In cases where the experimental input has a
systematic error dominated by theoretical uncertainties, we should
extract the latter with a flat distribution~\cite{Ciuchini:2000de}. We
have instead combined the errors in quadrature and extracted all the
experimental quantities with gaussian distributions. Within the
present accuracy, and taking into account the unknown non-perturbative
parameters, this procedure is fully justified. We have also verified
that by extracting the theoretical errors with a gaussian
distribution, we obtain very similar results. For more details on the
likelihood procedure, the reader is referred to~\cite{Ciuchini:2000de}, 
where all aspects are discussed at length.
\subsubsection*{Results with factorization}
We start by considering the case in which we use factorization and
take the CKM parameters $\vert V_{ub}\vert $ and $\gamma$ from other
experimental determinations.  We discuss first $BR(B^{+} \to
\pi^{+}\pi^{0})$ since in this case, due to isospin symmetry, we do
not have the complications due to penguin contractions.  Thus, at
fixed $\vert V_{ub}\vert $, the prediction for $BR(B^{+} \to
\pi^{+}\pi^{0})$ only depends on $f_\pi(0)$ (trivial dependences as
from $f_\pi$ will be omitted in this discussion).  By using the
theoretical estimate and uncertainty of $f_\pi(0)$
from~\cite{latticeff}, and taking into account the uncertainties on
$\vert V_{ub}\vert$, we predict in this case $BR(B^{+} \to
\pi^{+}\pi^{0})=(5.0 \pm 1.5) \times 10^{-6}$ in very good agreement
with the experimental average given in table~\ref{tab:inputs}.  A
complementary exercise is to use as input $\vert V_{ub}\vert$ and the
experimental value of $BR(B^{+} \to \pi^{+}\pi^{0})$ in order to
extract the value of $f_\pi(0)$.  In this case we find $f_\pi(0)= 0.28
\pm 0.06$, in very good agreement with lattice and QCD sum rules
estimates. This exercise shows that we do not need to rely on
theoretical calculations for the form factors. Indeed also for
$f_K(0)$ we only need $f_K(0)/f_\pi(0)$ which cannot differ too much
from one. Moreover it is likely that a large part of the uncertainties
of the theoretical predictions cancel in this ratio.

Here and in all the other cases where $\vert V_{ub}\vert $ and
$\gamma$ are taken from other experimental determinations, we use as
equivalent input parameters the values of $\bar \rho$ and $\bar \eta$
given in table~\ref{tab:inputs} from the UTA analysis of
ref.~\cite{Ciuchini:2000de}. These values correspond to
\begin{equation} \gamma = (54.8 \pm 6.2)^{0} \label{eq:gamma} \, .
\end{equation} 
By using $f_\pi(0)$ either from theory or from the fit to $BR(B^{+}
\to \pi^{+}\pi^{0})$ and assuming factorization, we then predict
$BR(B_{d} \to \pi^{+}\pi^{-})$ as a function of $\gamma$ only.
Besides, in order to analyze all $B \to K \pi$ decays, we only need
$f_K(0)/f_\pi(0)$ to which the previous considerations apply.
Alternatively we may take only $\vert V_{ub}\vert $ from the
experiments and fit the value of $\gamma$.  In the first case, the
results are given in table~\ref{tab:one} labeled as ``$\gamma$ UTA''
and show a generalized disagreement between predictions and
experimental data.  In the second case, the value of $\gamma$ is
fitted and the results are labeled as ``$\gamma$ free''. In this case
the disagreement is reduced for $BR(B^+ \to K^+ \pi^0)$ and $BR(B_d
\to K^+ \pi^-) $, and also for $BR(B_d \to \pi^+ \pi^-) $, but it
remains sizable for $BR(B_d \to K^0 \pi^0)$ and $BR(B^+ \to K^0
\pi^+)$.  The pattern $BR(B^+ \to K^0 \pi^+)$:$BR(B_d \to K^+ \pi^-)
$:$BR(B_d \to K^0 \pi^0)$:$BR(B^+ \to K^+ \pi^0)$=2:2:1:1, which is
suggested by the data, and is well reproduced when the contribution of
the charming penguins is large, as discussed in the following, is lost
in this case.  Moreover the fitted value of $\gamma=(163\pm 12)^{0}$
is in striking disagreement with the results of the UTA. 
Although one may question on the quoted uncertainty of the UTA result,
it is clearly impossible to reconcile the two numbers.  Thus either
there is new physics or $\Lambda_{QCD}/m_{b}$ corrections are
important. We now discuss the latter possibility.
\begin{table}
\begin{center}
\begin{tabular}{|c|c|c|c|}
\hline
$f_\pi(0)$& $0.27 \pm 0.08$ &
$f_K(0)/f_\pi(0)$ & $1.2 \pm 0.1$ \\ \hline
$\rho $ & $0.224 \pm 0.038$ & $\eta$ & $0.317 \pm 0.040 $ \\ \hline
$BR(B_d \to K^0 \pi^0)$ & $10.4 \pm 2.6 $ &  
$BR(B^+ \to K^+ \pi^0)$ & $12.1 \pm 1.7 $ \\ \hline
$BR(B^+ \to K^0 \pi^+)$ & $17.2 \pm 2.6$ &  
$BR(B_d \to K^+ \pi^-) $& $17.2 \pm 1.6$ \\ \hline
$BR(B_d \to \pi^+ \pi^-) $& $4.4 \pm 0.9 $ &  
$BR(B^+ \to \pi^+ \pi^0) $& $5.2 \pm 1.7$ \\
\hline
\end{tabular}
\end{center}
\caption{{\it Input values used in the numerical analysis. The form
    factors are taken from  
refs.~\cite{latticeff,qcdsrff}, the CKM parameters from
ref.~\cite{Ciuchini:2000de} and the BRs  
correspond to our average of CLEO, BaBar and Belle results 
\cite{cleobr,babarbr,bellebr}.  All the $BR$s are given in units of
$10^{-6}$.}} 
\label{tab:inputs}
\end{table}
\begin{table} 
 \begin{center} 
 \begin{tabular}{|c|c|c|c|c|c|}  \hline 
 $BR$ & $\gamma$ UTA & $\gamma$ free & $BR$ & $\gamma$ UTA & $\gamma$ free  \\ 
 \hline 
 $K^0 \pi^0$ & $5.9 \pm 0.2 $& $5.7 \pm 0.4 $& 
$K^+ \pi^0$ & $4.8 \pm 0.2 $&$ 9.1 \pm 0.5$\\ 
 $K^0 \pi^+$ & $11.7 \pm 0.5 $&$ 11.6 \pm 0.8 $& 
$K^+ \pi^-$ & $9.8 \pm 0.4 $&$ 17.7 \pm 1.0$\\ 
 $ \pi^+ \pi^-$ & $8.5 \pm 0.3 $&$ 5.1 \pm 0.7 $& 
$\pi^+ \pi^0$ & $4.2 \pm 0.2 $& $5.4 \pm 0.6$\\ 
 $ \pi^0 \pi^0$ & $0.19 \pm 0.01 $&$ 0.59 \pm 0.04 $& 
&&\\ 
\hline  
 \end{tabular} 
 \end{center} 
 \caption{{\it Results for the $BR$s obtained with factorization 
without charming or GIM penguins. All the $BR$s are given in units of
$10^{-6}$.}}  
 \label{tab:one} 
\end{table} 
\subsubsection*{Factorization with Charming and GIM penguins}
\begin{figure}
\center
\begin{tabular}{cc}
\epsfig{figure=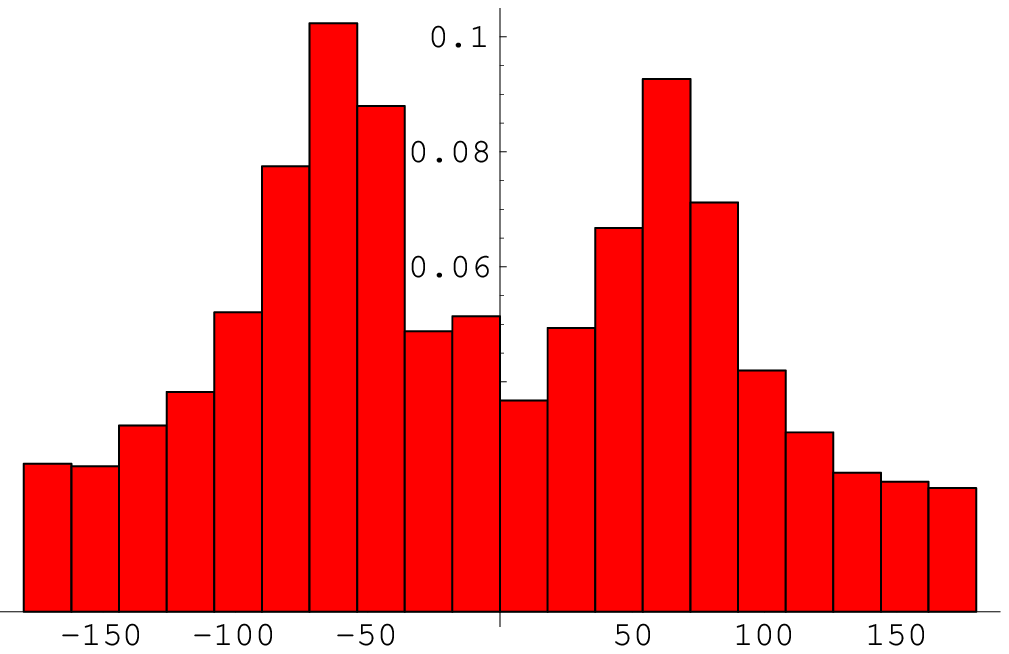,height=1.7in} & \epsfig{figure=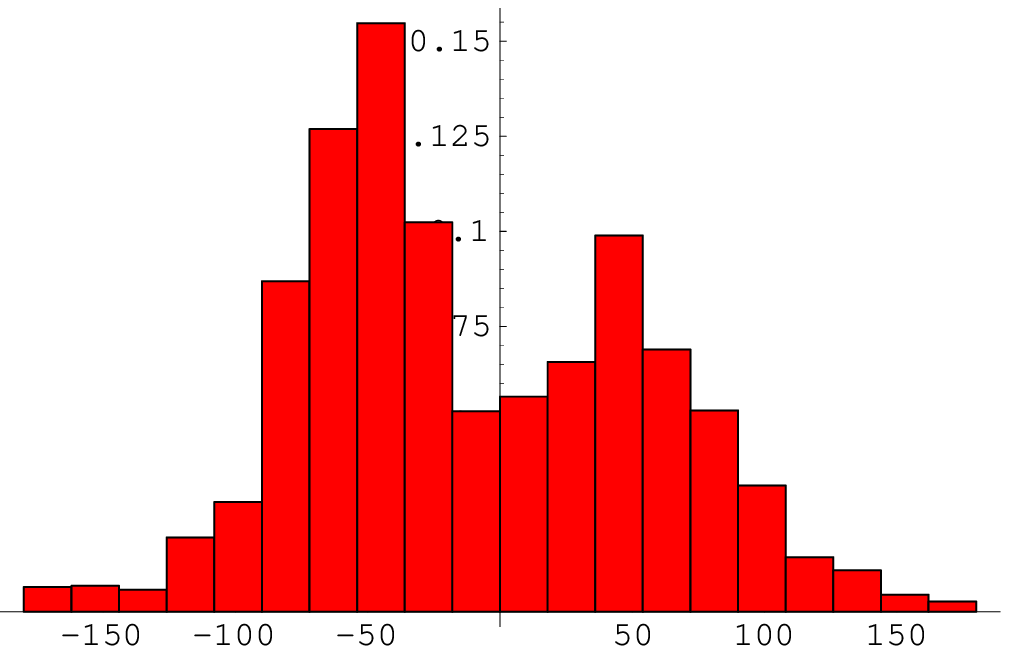,height=1.7in}\\
\end{tabular}
\caption{{\it p.d.f. for $\phi $, in the case where only $\tilde P_1$
    (left) and both $\tilde P_1$  
and $\tilde P_1^{\sss {\rm GIM}}$ (right) are included.}}
\label{fig:phi}
\end{figure}

We now discuss the effects of charming penguins, parameterized by
$\tilde P_{1}$.  $\tilde P_{1}$ is a complex amplitude that we fit on
the $B \to K \pi$ $BR$s.  In order to have a reference scale for its
size, we introduce a suitable ``Bag'' parameter, $\tilde B_{1}$, by
writing 
\begin{equation} 
\tilde P_{1} = \frac{G_{F}}{\sqrt{2}} \,
  f_{\pi} \, f_{\pi}(0) \, g_{1} \tilde B_{1} \, , 
\end{equation}
where $G_{F}$ is the Fermi constant.  We use $f_{\pi}(0)$ for both $B \to
K \pi$ and $B \to \pi\pi$ channels since, as mentioned before, for
charming penguins we work in the $SU(3)$ limit.  $g_{1}$ is a
Clebsh-Gordan parameter depending on the final $K \pi$ ($\pi\pi$)
channel.  In the case where $\vert V_{ub}\vert $ and $\gamma$ are
taken from the UTA, by fitting the $B \to K \pi$ channels and $B^+ \to
\pi^+ \pi^0$ only, we find
\begin{equation} 
\vert \tilde B_{1} \vert =
  0.14 \pm 0.05 \, . 
\end{equation} 
Note that the size of the charming penguin effects is of the
expected magnitude.  As for the phase $\phi=Arg(\tilde B_{1})$, it is very
instructive to consider its distribution, which is displayed in
fig.~\ref{fig:phi}: the preferred value of $\phi$ has a sign
ambiguity since we are fitting the average of the $B^{0}_{d}$ and
$\bar B^{0}_{d}$ $BR$s (or of the $B^{+}$ and $B^{-}$ $BR$s). The
ambiguity can be resolved by measuring separately particle and
anti-particle $BR$s.
\begin{figure}
\center
\begin{tabular}{cc}
\epsfig{figure=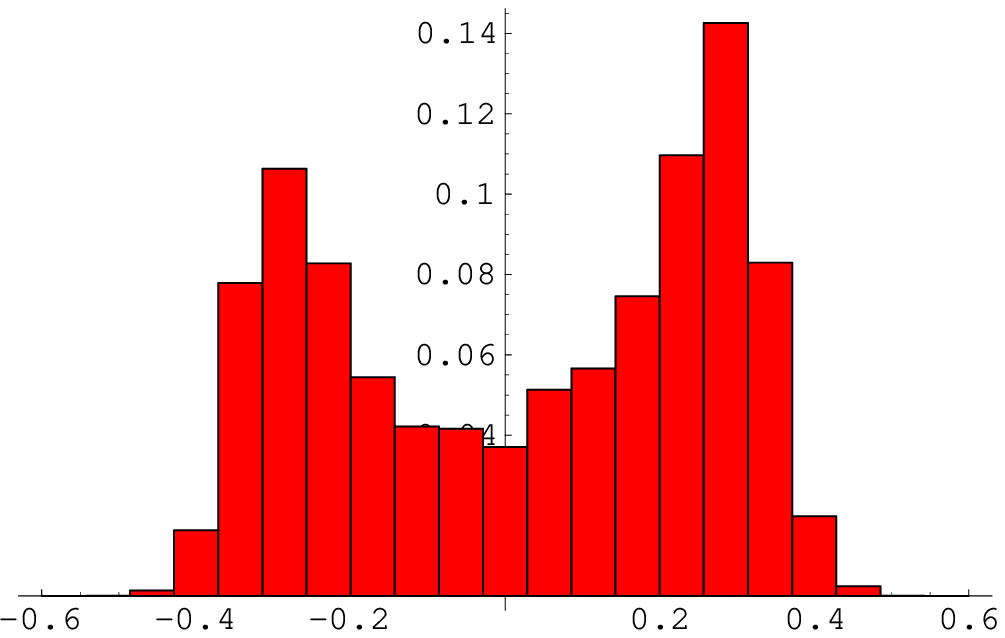,height=1.6in} &
\epsfig{figure=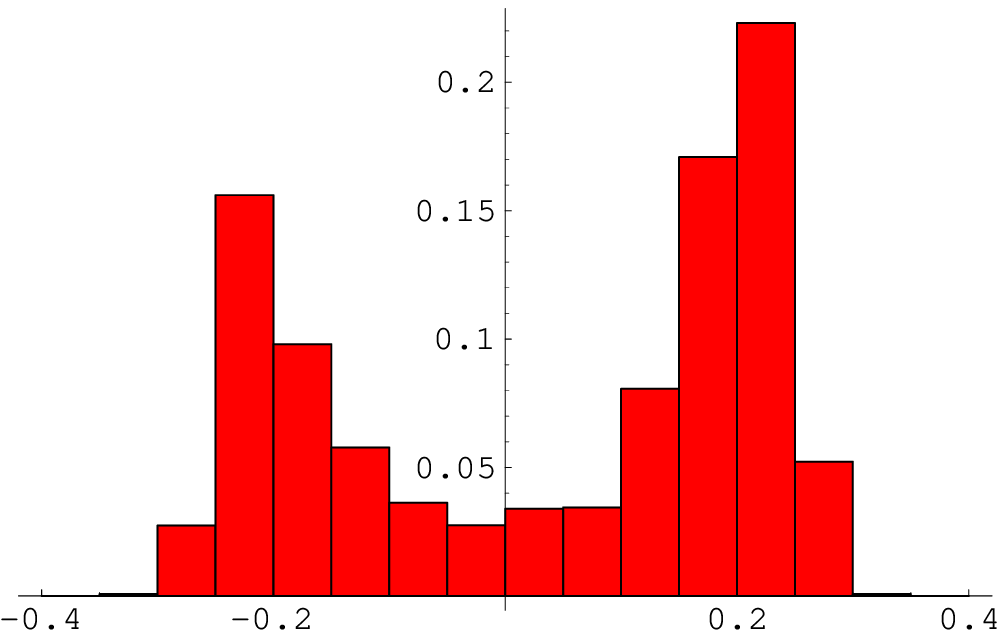,height=1.6in}\\ 
\epsfig{figure=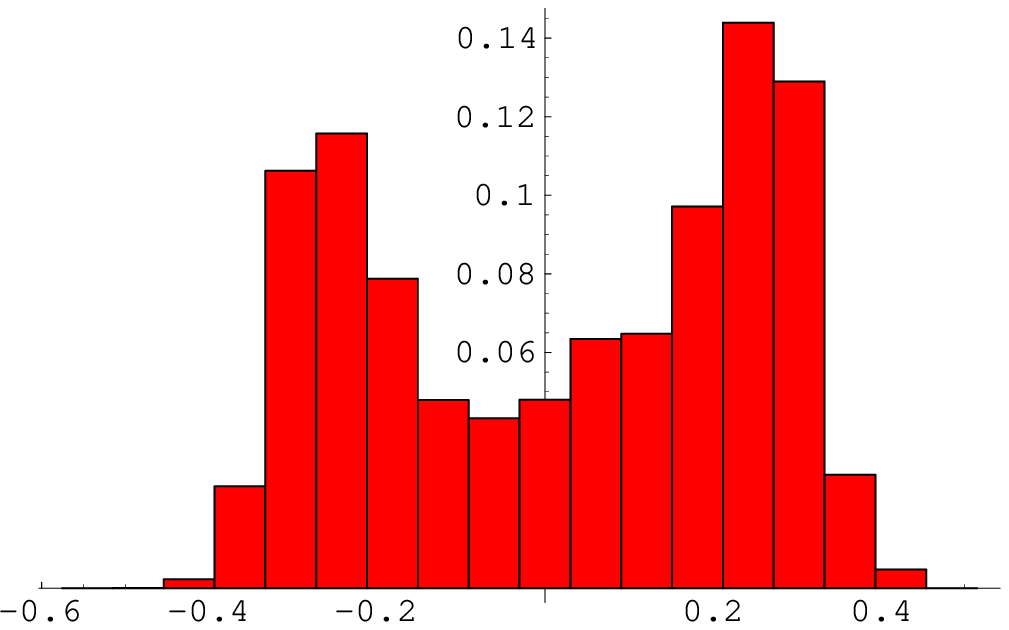,height=1.6in} &
\epsfig{figure=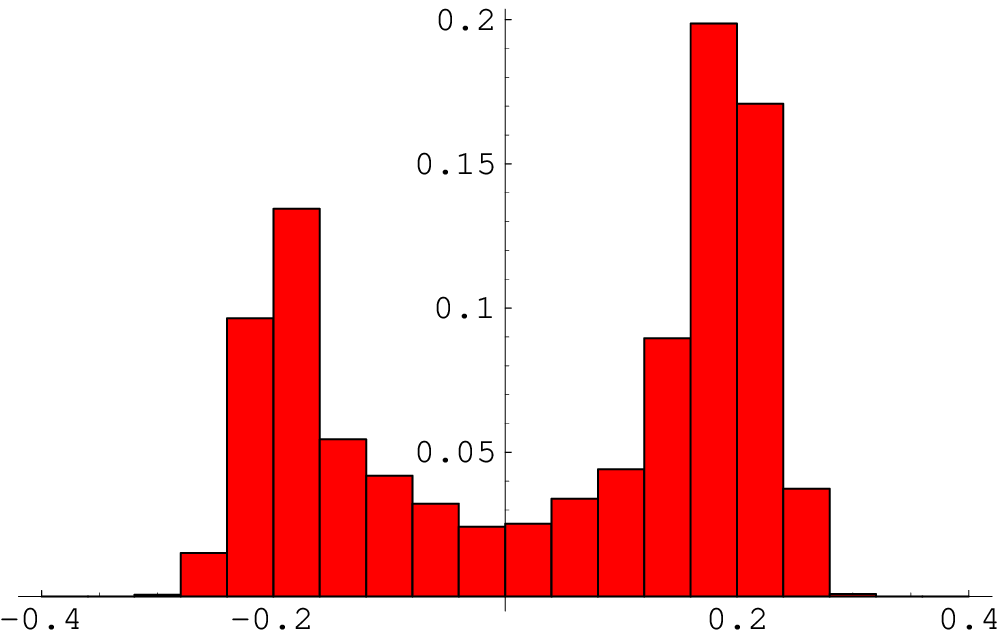,height=1.6in}\\ 
\epsfig{figure=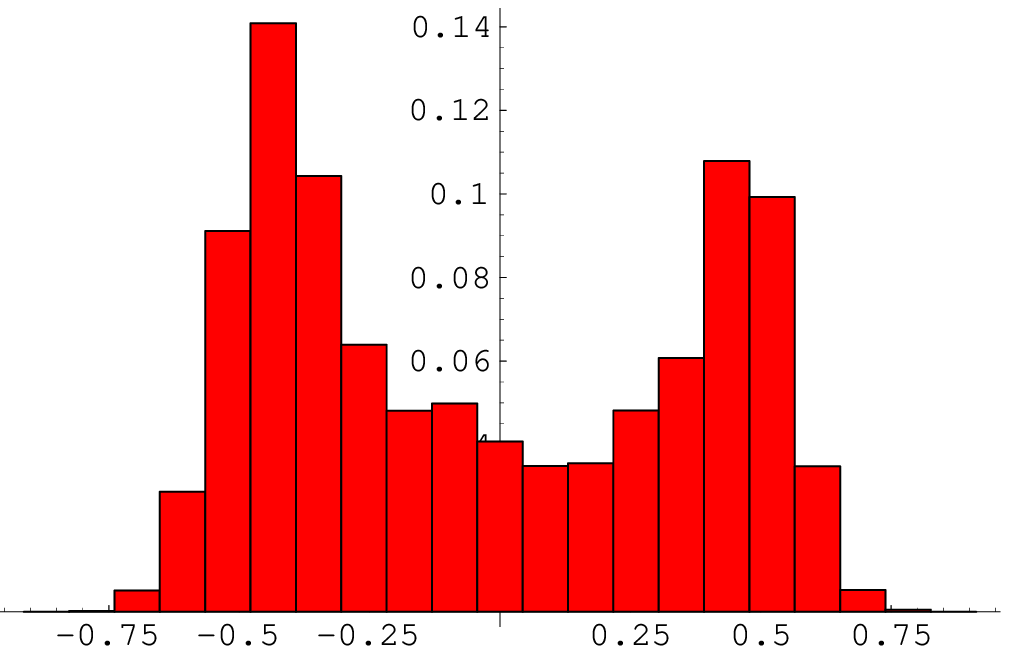,height=1.6in} &
\epsfig{figure=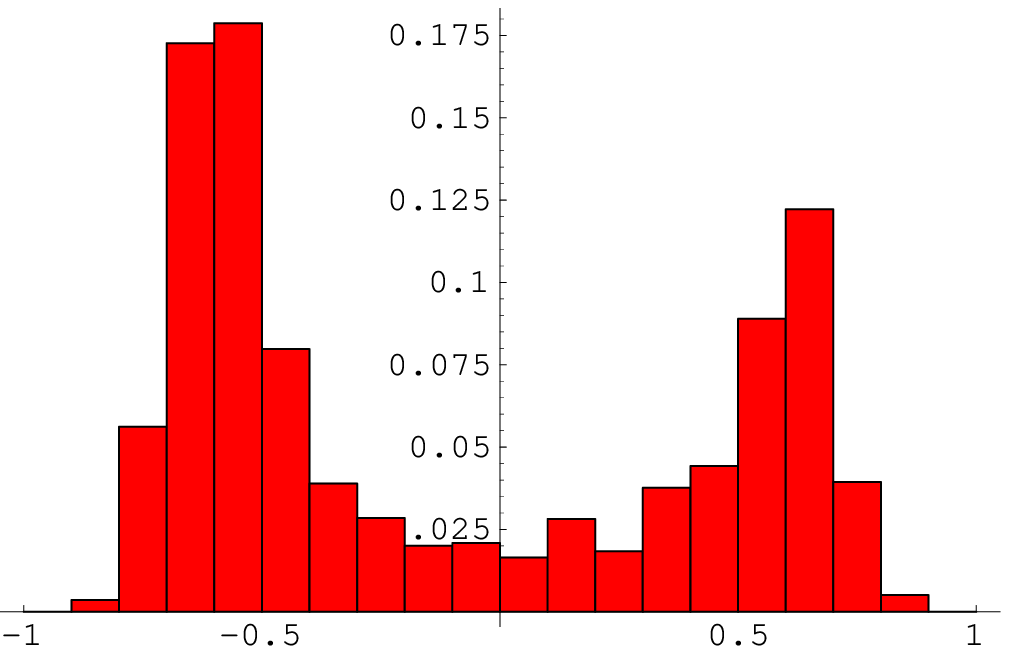,height=1.6in}\\ 
\end{tabular}
\caption{{\it p.d.f. for the largest  CP asymmetries, in the case
    where only $\tilde P_1$ (left) and  
both $\tilde P_1$ 
and $P_1^{\sss {\rm GIM}}$ (right) are included.
From top to bottom, we give ${\cal A}(B^+ \to K^+ \pi^0)$, 
${\cal A}(B_d \to K^+ \pi^-)$ and ${\cal A}(B_d \to \pi^+ \pi^-)$. }}
\label{fig:asy}
\end{figure}
By using the distribution on the left of fig.~\ref{fig:phi}, we
compute the mean value of $\vert \phi\vert$ with the result $\vert
\phi\vert = (75 \pm 44)^{o}$ and leave the sign undetermined. This is a
reasonable procedure, given the approximate symmetry of the distribution and
the large uncertainty.  In view of the discussion of the
particle-antiparticle asymmetry which we present at the end of this
paper, we note here that the value of $\phi$ could be rather large.
In table~\ref{tab:two} we give the corresponding predicted values and
uncertainties for the relevant branching ratios (label ``Charming'').  We
observe a remarkable improvement for the $K \pi$ channels and a large
shift in the value of $BR(B_d \to \pi^0 \pi^0)$~\footnote{ This effect was already noticed 
in~\cite{charming}.}, in spite of the fact
that in the latter case penguin effects are not Cabibbo enhanced (the
$\pi^{0} \pi^{0}$ amplitude is however colour
suppressed).  The predicted value for $BR(B_d \to
\pi^+ \pi^-) $ remains however much larger than the experimental one.

If one fits the $B \to K \pi$ channels, $B^{+} \to \pi^+ \pi^0$ and
$B_d \to \pi^+ \pi^-$ simultaneously, one finds a better agreement for
$BR(B_d \to\pi^+ \pi^-)$ but a rather small value for $BR(B^{+} \to
\pi^+ \pi^0)$ (column ``Charming with $\pi^{+ } \pi^{-}$'' of
table~\ref{tab:two}).  This happens at the price of reducing the
fitted value of the form factor, $f_{\pi}(0)\sim 0.22$, which is
pushed down by $BR(B_d \to\pi^+ \pi^-)$.  In fact the latter has an
experimental error much smaller than $BR(B^{+} \to \pi^+ \pi^0)$, and
therefore governs the fit.  However, we do not think that this is the
correct procedure: theoretically, $BR(B^{+} \to \pi^+ \pi^0)$ is on
much more solid grounds than $BR(B_d \to\pi^+ \pi^-)$, since it is not
affected by penguins or annihilations, and thus is much more suitable
to constrain $f_{\pi}(0)$.
\begin{table} 
\begin{center} 
\begin{tabular}{|c|c|c|c|c|c|c|c|}
\hline $BR$ & Charming & Charming & Charming & $BR$ & Charming &
Charming & Charming \\ & & with $\pi^{+ } \pi^{-}$ & + GIM & & & with
$\pi^{+ } \pi^{-}$ & + GIM \\ \hline $K^0 \pi^0$ & $9.2 \pm 1.1 $&$8.7
\pm 0.9$& $8.8 \pm 1.0 $& $K^+ \pi^0$ & $9.2 \pm 0.8 $&$9.3 \pm 0.7$&$
9.3 \pm 0.7$\\ $K^0 \pi^+$ & $18.3 \pm 2.1 $&$17.4 \pm 1.8$&$ 17.6 \pm
1.8 $& $K^+ \pi^-$ & $18.2 \pm 1.4 $&$18.6\pm 1.4$&$ 18.4 \pm 1.3$\\ $
\pi^+ \pi^-$ & $9.1 \pm 2.5 $&$5.1\pm 1.8$ &$ 4.7 \pm 0.8 $& $\pi^+
\pi^0$ & $4.8 \pm 1.4 $&$2.7 \pm 0.5$ &$3.5 \pm 0.9$\\ $ \pi^0 \pi^0$
& $0.37 \pm 0.05 $&$0.36 \pm 0.05$&$ 0.69 \pm 0.30 $& && &\\ \hline
\end{tabular} 
\end{center} 
\caption{{\it $BR$s with charming or
      charming and GIM penguins.  All the $BR$s are given in units of
      $10^{-6}$.}}  
\label{tab:two}
\end{table} 

In order  to reduce the predicted $BR(B_d \to
\pi^+ \pi^-) $ without affecting $BR(B^{+}  \to \pi^+ \pi^0)$, 
one may  include other effects  of the same order of the 
charming penguins, as for example the GIM  penguins introduced in 
ref.~\cite{charming}.  In this case we fit all the $BR$s given in
table~\ref{tab:inputs}. With GIM and charming
penguins included, we find
\begin{eqnarray} \vert \tilde B_{1} \vert &=& 0.16 \pm 0.03\, , \quad \quad  
\vert \phi \vert= (56 \pm 32)^o \, , 
\nonumber \\
\vert \tilde B^{\sss {\rm GIM}}_{1} \vert &=& 0.23 \pm 0.11\, , \quad \quad  
\vert \phi^{\sss {\rm GIM}} \vert = (135 \pm 37)^o \, , 
\end{eqnarray}
where the notation is self-explaining.  We have given the absolute
value of $\phi$ since, as in the previous case, the sign ambiguity
persists when we include GIM penguins. The distribution is also shown
in fig.~\ref{fig:phi}.  The results for the $BR$s can be found in
table~\ref{tab:two} with the label ``Charming+GIM".  They show that
the extra GIM parameter improves the agreement for the measured $B \to
\pi \pi $ $BR$s.  We do not claim, however, to be able to predict
$BR(B_d \to \pi^+ \pi^-) $: our results instead show that accurate
predictions for $B_{d} \to \pi \pi$ decays can only be obtained by
controlling quantitatively the ${\cal O}(\Lambda_{QCD}/m_{b})$
corrections, which is presently beyond the theoretical reach.
Estimates for charming penguin effects can also be obtained by using some
phenomenological model, as for example done in ref.~\cite{pham}. We
 observe that the sensitivity of the $BR$s to the value of
$\gamma$ is lost, with the present experimental accuracy, once penguin
effects are introduced. Indeed when one tries to fit $B_{1}$
($B_{1}^{\sss {\rm GIM}}$) and $\gamma$ simultaneously, one finds that
the value of $\gamma$ is essentially undetermined.  From the above
discussion it clearly emerges that one of the important step for
the improvement of this kind of analyses is a more precise measurement
of $BR(B^{+} \to \pi^+ \pi^0)$.
\subsubsection*{Particle--Antiparticle asymmetries for the Branching 
  Ratios} The large absolute values of $\phi$, and the sizable
effects that penguins have on the $BR$s, stimulated us to consider
whether we could find observable particle-antiparticle asymmetries as
the one defined in eq.~(\ref{eq:asy}). We find large effects in
$BR(B^{+} \to K^+ \pi^0) $, $BR(B_d \to K^+ \pi^-) $ and $BR(B_d \to
\pi^+ \pi^-)$, as shown in fig.~\ref{fig:asy}.  As discussed before,
for $BR(B_d \to
\pi^+ \pi^-)$ our predictions suffer from very large uncertainties due
to contributions which cannot be fixed theoretically. For this reason,
the values of the asymmetry  reported in table~\ref{tab:asy} are only
an indication that a large asymmetry could be observed also in this channel.
The sign ambiguity
of $\phi$ is reflected in  the
asymmetry ${\cal A}\sim \sin \gamma \sin \phi$. This ambiguity can be
solved only by an experimental measurement or, but this is extremely
remote, by a theoretical calculation of the relevant amplitudes. For
each channel, we give the absolute value of the asymmetry in
table~\ref{tab:asy}.  Note that within factorization all asymmetries
would be unobservably small, since the strong phase is a perturbative
effect of ${\cal O}(\alpha_{s})$~\cite{beneke}.  The possibility of
observing large asymmetries in these decays opens new perspectives. 
These points will be the subject of a future study.
 \begin{table} 
 \begin{center} 
 \begin{tabular}{|c|c|c|c|c|c|}  \hline 
 $\vert {\cal A} \vert$ & Charming & Charming + GIM &  $\vert {\cal
 A}\vert$ & Charming & Charming + GIM  \\  
 \hline 
 $K^0 \pi^0$ & $0.02 \pm 0.01 $& $0.05 \pm 0.03 $& 
$K^+ \pi^0$ & $0.23 \pm 0.10 $&$ 0.18 \pm 0.06$\\ 
 $K^0 \pi^+$ & $0.00 \pm 0.00 $&$ 0.03 \pm 0.03 $& 
$K^+ \pi^-$ & $0.21 \pm 0.10 $&$ 0.17 \pm 0.06$\\ 
 $ \pi^+ \pi^-$ & $0.36 \pm 0.16 $&$ 0.52 \pm 0.18 $& 
 $ \pi^0 \pi^0$ & $0.40 \pm 0.19 $&$ 0.58 \pm 0.29 $\\ 
\hline  
 \end{tabular} 
 \end{center} 
 \caption{{\it Absolute values of the rate CP asymmetries for $B \to K
 \pi$ and  
 $B \to \pi \pi$ decays. The columns labeled by ``Charming'' and
 ``Charming + GIM''  
 correspond respectively to the
 cases in which only $\tilde P_1$ and both $\tilde P_1$ and 
 $\tilde P_1^{\sss {\rm GIM}}$ are introduced. The asymmetry in 
 $B \to \pi^{+} \pi^{0}$ vanishes exactly.}} 
 \label{tab:asy} 
\end{table} 

\section*{Conclusion}
We have analyzed the predictions of factorization for $B \to \pi\pi$
and $B \to K \pi$ decays. We note that the normalization of all the
other $BR$s is essentially fixed by the value of $BR(B^{+} \to \pi^{+}
\pi^{0})$ and $SU(3)$ symmetry.  Even taking into account the
uncertainties of the input parameters, we find that factorization is
unable to reproduce the observed $BR$s. The introduction of charming
and GIM penguins~\cite{charming} allows to reconcile the theoretical
predictions with the data.  It also shows however that it is not
possible, with the present theoretical and experimental accuracy, to
determine the CP violation angle $\gamma$.  Contrary to factorization,
we predict large asymmetries for several of the particle--antiparticle
$BR$s, in particular $BR(B^{+} \to K^+ \pi^0) $, $BR(B_d \to K^+
\pi^-) $ and $BR(B_d \to \pi^+ \pi^-)$. This opens new perspectives
for the study of CP violation in $B$ systems.
\section*{Acknowledgments}
We thank G. Buchalla and C. Sachrajda for useful discussions on our work.  M.C.
thanks the TH division at CERN where part of this work has been done. 
L.S. thanks Hsiang-nan Li for very informative discussions and J. Matias 
for pointing out a misprint.

\end{document}